\documentclass[12pt]{article}
\usepackage{graphicx}
\usepackage{amssymb}
\usepackage{hyperref}
\usepackage{float}
\usepackage{cite}

\newcommand\pubnumber{}
\newcommand\pubdate{\today}

\def\institute{DAMTP, University of Cambridge, Wilberforce Road, Cambridge, CB3 0WA,
United Kingdom}
\def\support{\footnote{This work was supported by the European Research
Council under the European Union’s Horizon 2020 research and innovation Programme (grant agreement
n.950246) and in part by STFC consolidated grant ST/T000694/1.}}

\def\Title#1{\begin{center} {\Large #1 } \end{center}}
\def\Author#1{\begin{center}{ \sc #1} \end{center}}
\def\Address#1{\begin{center}{ \it #1} \end{center}}

\newcommand\pubblock{\rightline{\begin{tabular}{l} \pubnumber\\
         \pubdate  \end{tabular}}}
\newenvironment{Abstract}{\begin{quotation}  }{\end{quotation}}
\newenvironment{Presented}{\begin{quotation} \begin{center} 
             PRESENTED AT\end{center}\bigskip 
      \begin{center}\begin{large}}{\end{large}\end{center} \end{quotation}}
\def\Acknowledgements{\bigskip  \bigskip \begin{center} \begin{large}
             \bf ACKNOWLEDGEMENTS \end{large}\end{center}}

\def\beq{\begin{equation}}
\def\eeq#1{\label{#1}\end{equation}}
\def\eeqn{\end{equation}}

\def\beqa{\begin{eqnarray}}
\def\eeqa#1{\label{#1}\end{eqnarray}}
\def\eeqan{\end{eqnarray}}

\let\bar=\overbar

\def\Dslash{\not{\hbox{\kern-4pt $D$}}}
\def\dslash{\not{\hbox{\kern-2pt $\del$}}}

\def\msb{{\bar{\ssstyle M \kern -1pt S}}}

\begin{document}
\begin{titlepage}
\pubblock

\vfill
\Title{Top, Higgs, Diboson and Electroweak Fit to the Standard Model Effective Field Theory}
\vfill
\Author{ Maeve Madigan\support}
\Address{\institute}
\vfill
\begin{Abstract}
	The Standard Model Effective Field Theory (SMEFT) provides a powerful theoretical framework for searching for subtle deviations from the Standard Model.
In this talk, we present the results from a global fit of the dimension-6 operators of the SMEFT to a combination of Higgs, top, diboson and electroweak precision observables.  
SMEFT constraints driven by data from the top quark sector are
	highlighted. We explore the interplay between the top and Higgs sectors,
	emphasising the need for a global approach to constraining the SMEFT.
\end{Abstract}
\vfill
\begin{Presented}
$15^\mathrm{th}$ International Workshop on Top Quark Physics\\
Durham, UK, 4--9 September, 2022
\end{Presented}
\vfill
\end{titlepage}
\def\thefootnote{\fnsymbol{footnote}}
\setcounter{footnote}{0}

\section{Introduction}
Experimental probes of new physics beyond the Standard Model (BSM) have been taken to an unprecedented level by data from the Large Hadron Collider (LHC).
In particular Run II has provided many new precision measurements, for example differential measurements of top quark pair production~\cite{Sirunyan:2018wem,ATLAS:2019czt}.
Although there is no direct evidence for new physics within this data, the abundance of precision and high energy information
provides an excellent opportunity for performing
indirect searches for new physics, making us of the framework of the Standard Model Effective Field Theory (SMEFT) in this endeavour~\cite{Ellis:2020unq,Ethier:2021bye,Ellis:2018gqa,Biekotter:2018rhp,Brivio:2019ius,Hartland:2019bjb,Buckley:2015lku}.
In this talk, we present the results of a global analysis of
data from the top, Higgs, diboson and electroweak precision sectors~\cite{Ellis:2020unq}.

The global fit is performed using the {\tt Fitmaker} code\footnote{available from the following {\tt Gitlab} link:~\href{https://gitlab.com/kenmimasu/fitrepo}{https://gitlab.com/kenmimasu/fitrepo}}, employing the method
of least-squares to constrain the dimension-6 operators of the Warsaw basis~\cite{Grzadkowski:2010es}.  
A `top-specific' flavour symmetry is applied to the dimension-6 operators~\cite{AguilarSaavedra:2018nen},
reducing them to a set of 34 operators.
The symmetry, denoted by
$SU(2)_{q} \times SU(2)_{u} \times SU(3)_{d} \times SU(3)_{l} \times SU(3)_{e}$,
relaxes the assumption of flavour universality for operators involving the top quark.
Under this symmetry four-fermion operators involving the top quark are allowed, as well as chirality-flipping interactions
such as $O_{tG}$, $O_{tB}$ and $O_{tH}$, for example.
Theory predictions in the SMEFT are calculated at linear order in the dimension-6 operators and
to leading order in QCD
using {\tt SMEFTsim}~\cite{Brivio:2017btx} and {\tt SMEFT@NLO}~\cite{Degrande:2020evl}.  
These operators are fit to a total of
341 datapoints,
including Simplified Template Cross Section  measurements of the Higgs sector~\cite{Aad:2019mbh},
top pair production invariant mass distributions~\cite{Sirunyan:2018wem},
top pair production charge asymmetry measurements~\cite{ATLAS:2019czt}
and measurements of the W boson polarisation in top quark decays~\cite{Aad:2020jvx}.
For more details of the data and the technical settings, we refer to our main work, Ref.~\cite{Ellis:2020unq}.

\begin{figure}[h]
\centering
\includegraphics[width=13cm]{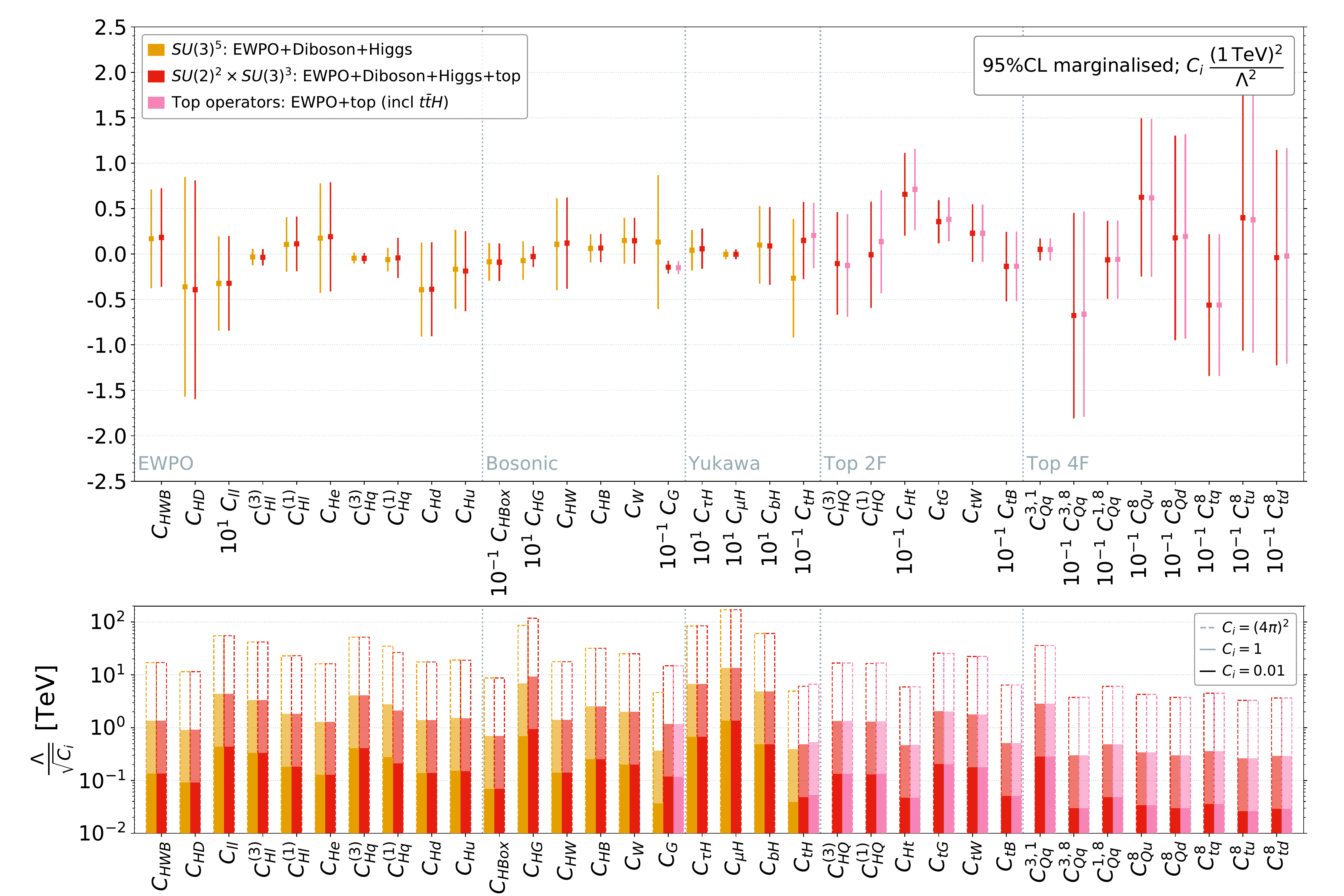}
\caption{Global fit to Higgs, top, diboson and electroweak precision data.
	Constraints at 95\% CL are shown on each
	coefficient in the top-specific flavour scenario
	resulting from a marginalisation over the remaining operators in the fit.
	The impact of removing top data, and of removing Higgs and diboson data, is shown
	by the fits in orange and pink respectively.
	}
	\label{fig:marg}
\end{figure}

\section{Global SMEFT fit}
Figure~\ref{fig:marg} shows the constraints at 95\% CL on each of the 34 dimension-6 operators in the top-specific flavour symmetry, resulting from a fit
to the combined set of top, Higgs, diboson and electroweak precision data. 
The constraints shown on each operator in the upper panel result from a marginalisation over all other operators in the fit.
The lower panel indicates the corresponding lower limits at 95\% CL on the new physics scale $\Lambda$, taking
coefficients $C_{i}=0.01$, $C_{i}=1$ and $C_{i}=4 \pi$ as benchmark values.

On the right hand side, we show in pink the impact on the top-specific operators
of removing data from the Higgs and diboson sectors. The majority of the constraints are
stable under the removal of the Higgs and diboson data, although small changes
are observed in the constraints on the coefficients $C_{tH}$ and $C_{tG}$, 
indicating that the inclusion of Higgs data is advantageous in constraining these operators.
Similarly, by comparing the red and orange constraints we see the impact
of removing data from the top quark sector.
In this case the constraints on $C_{HG}$, $C_{G}$ and $C_{tH}$ widen, indicating that 
these coefficients benefit from the inclusion of top quark data in the marginalised fit.
These subtle hints at an interplay between the top and Higgs sectors will be further explored in the next section.

\begin{figure}[h]
\centering
\includegraphics[height=10cm]{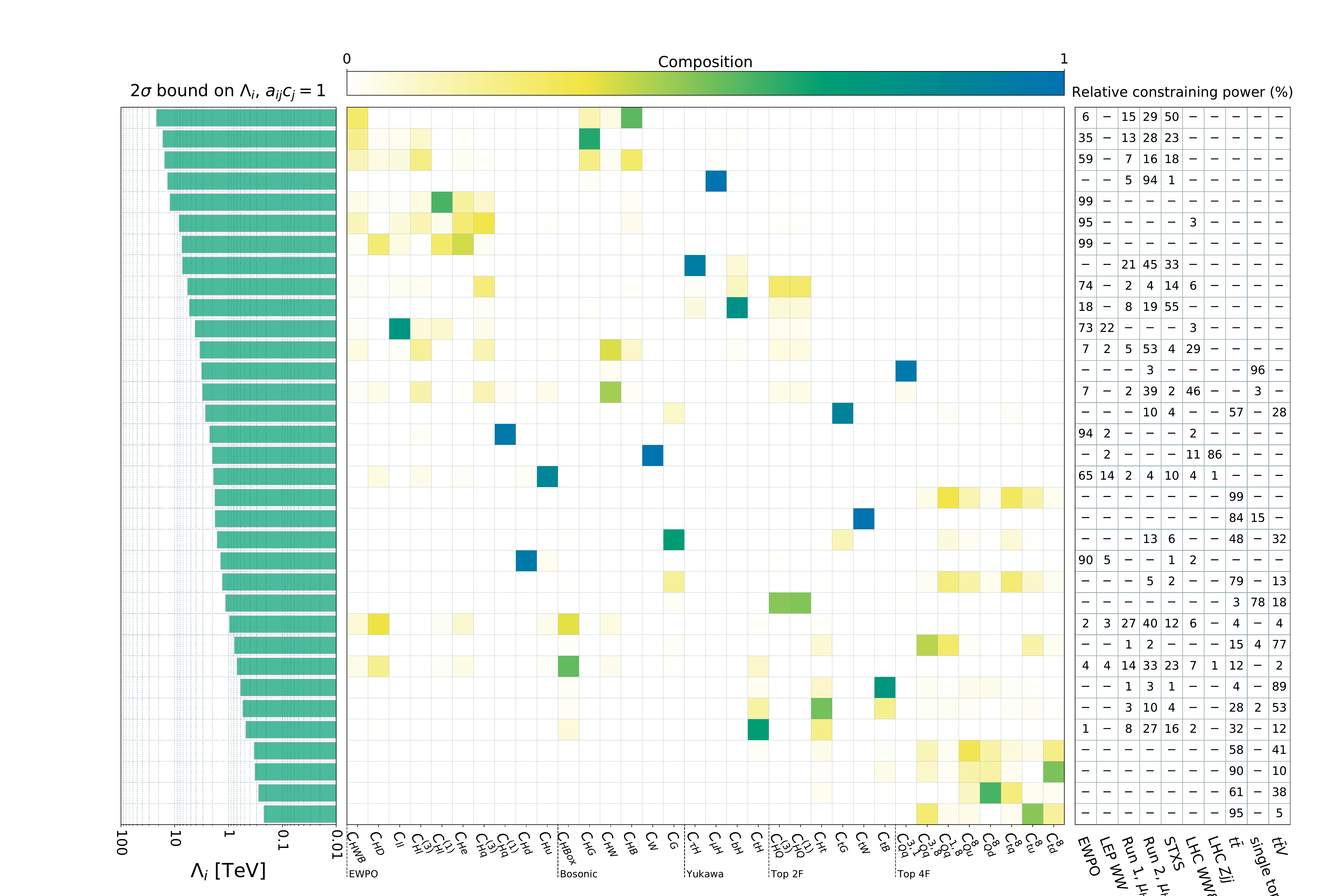}
\caption{Results from a principal component analysis of the global fit.
	Eigenvectors of the covariance matrix are denoted by rows in the central panel, and
	their SMEFT operator composition is indicated by the coloured squares.
	The right panel shows the relative constraining power of each of the datasets included in the global fit,
	while the left panel presents the lower limit at 95\% CL on the new physics scale $\Lambda$ associated to each eigenvector.}
\label{fig:eigensystem}
\end{figure}

The extent to which data from the top sector constrains the SMEFT coefficients can be seen from a principal component 
analysis (PCA) of the global fit, displayed in Figure~\ref{fig:eigensystem}.
Each row of the central panel represents an eigenvector, and the operator composition of each eigenvector is shown by 
the coloured squares.  
The lower limit at 95\% CL on the new physics scale $\Lambda$ associated to each eigenvector is 
given in the left panel,
while the right panel indicates the relative constraining power of each dataset.
We observe a number of strong constraints on the SMEFT originating from top quark data.
The eigenvector predominantly composed of the coefficient $C_{tG}$ is constrained to $\Lambda > 2$ TeV,
with top pair production and $t \bar{t} + V$ data providing the dominant constraining power.
Two rows above this, we see that the four-fermion operator $O_{Qq}^{1,3}$ is similarly constrained to $\Lambda > 2$ TeV, in this case by single-top data,
and a few rows below this
the $O_{tW}$ eigenvector
is constrained to $\Lambda > 1$ TeV by $W$ boson polarisation measurements in top quark decays.  
In contrast, in the lower right hand side of the PCA, we see that eigenvectors composed of the four-fermion operators are more weakly constrained by $t \bar{t}$ data to $\Lambda \gtrsim 300$ GeV.  In this case the validity of the SMEFT expansion is not guaranteed,
and we expect that the quadratic contributions from dimension-6 operators will become important,
as seen in~\cite{Ethier:2021bye}.
However, we emphasise that these constraints result from a marginalised fit in which all 34 coefficients are allowed to float.  Realistic BSM scenarios may generate a subset of these operators, resulting in a correspondingly higher new physics scale.

\section{Top-Higgs interplay}
\begin{figure}[h]
\centering
\includegraphics[height=7.5cm]{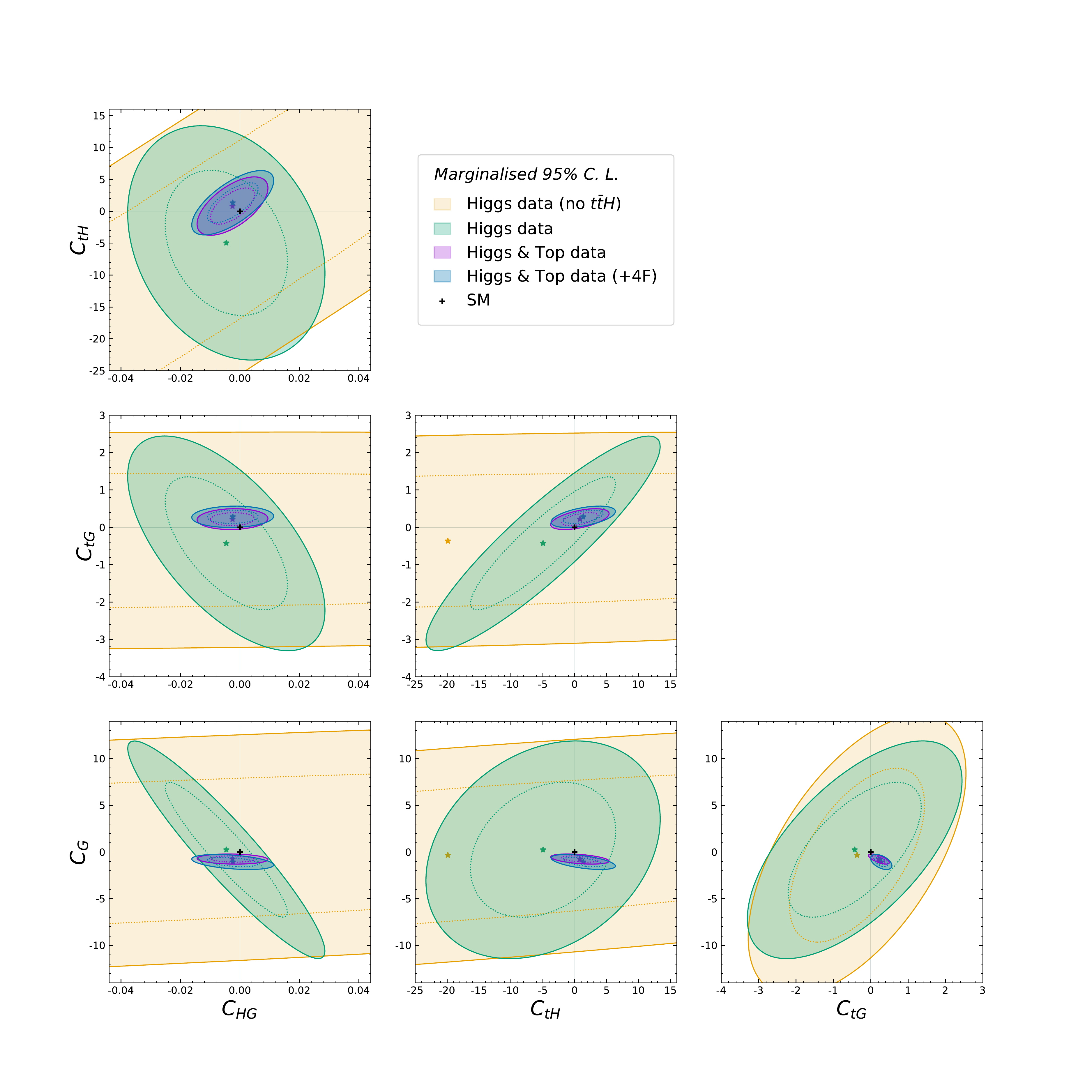}
	\caption{Constraints at 95\% CL resulting from a fit of the coefficients  $\{ C_{G}, C_{tG}, C_{H \Box}, C_{HG}, C_{HW}, C_{HB}, C_{tH}, C_{bH}, C_{\mu H}, C_{\tau H} \}$ to Higgs data excluding $t \bar{t} H$, Higgs data including $t \bar{t} H$, and Higgs \& top data.  In the latter case we assess the impact of switching on the coefficients of the four-fermion operators in the fit.}
\label{fig:tophiggs}
\end{figure}

Next, we further investigate the level of
interplay between top and Higgs data in the global fit.  
In Figure~\ref{fig:tophiggs} we display the results of a fit to the coefficients relevant to Higgs processes,
$\{ C_{H \Box}, C_{HG}, C_{HW}, C_{HB}, C_{tH}, C_{bH}, C_{\mu H}, C_{\tau H} \}$, as well as $C_{tG}$ and $C_{G}$.  
The latter
contribute to both top pair production as well as to the gluon gluon fusion mode of Higgs and Higgs + jet production.
Constraints at 95\% CL are displayed for the pairs of operator coefficients shown in each panel, obtained from a marginalisation over the remaining operators in the fit.  

By comparing the constraints obtained from Higgs data with and without the inclusion of $t\bar{t}H$ data, in green and yellow respectively, we see that a flat direction 
between $C_{tH}$ and $C_{HG}$ is removed by the inclusion of the measurement of the top Yukawa, improving the sensitivity to both $C_{tH}$ and $C_{HG}$ as a result.
A marked improvement in this sensitivity is seen when all top quark data is included in the fit, as shown by the comparison between green and purple ellipses.  
The top quark data is able to provide strong constraints on $C_{tG}$ and $C_{G}$,
allowing $C_{HG}$ to be further constrained by Higgs gluon gluon fusion production data.
The result is a vast decrease in the area of the ellipses allowed at 95\% CL, as well
as a supression of correlations between operators.
However, we note that top quark data is modified by SMEFT operators other than $C_{G}$ and $C_{tG}$, in particular
four-fermion operators, and that setting their contribution to zero 
may not be physically well-motivated in all BSM scenarios.
The blue constraints show the impact of allowing these coefficients to vary in the fit,
and we find that our conclusions are unchanged:
the constraints from the combination of Higgs and top data are improved relative to those from Higgs data alone.

\section{Conclusions}
Searches for new physics within the SMEFT framework benefit from the approach of a global fit.
In this talk we have presented the results of a global analysis of data from the Higgs, top, diboson 
and electroweak sectors, constraining 34 operators of the dimension-6 SMEFT.
We observe some strong constraints originating from the top sector, and the need for a global fit is emphasised by the observation of an interplay between the 
top and Higgs sectors.

\Acknowledgements
I am grateful to the organisers of the 
15th International Workshop on Top Quark Physics,
Durham,
for the opportunity to present
this research, and to John Ellis, Ken Mimasu, Veronica Sanz and Tevong You, with whom the work presented here was carried out.

\bibliographystyle{utphys}
\bibliography{references}

\providecommand{\href}[2]{#2}\begingroup\raggedright\begin{thebibliography}{10}

\bibitem{Sirunyan:2018wem}
{\bfseries CMS} Collaboration, A.~M. Sirunyan {\em et~al.}, ``{Measurement of
  differential cross sections for the production of top quark pairs and of
  additional jets in lepton+jets events from pp collisions at $\sqrt{s} =$ 13
  TeV},'' \href{http://dx.doi.org/10.1103/PhysRevD.97.112003}{{\em Phys. Rev.
  D} {\bfseries 97} no.~11, (2018) 112003},
  \href{http://arxiv.org/abs/1803.08856}{{\ttfamily arXiv:1803.08856
  [hep-ex]}}.

\bibitem{ATLAS:2019czt}
{\bfseries ATLAS} Collaboration, ``{Inclusive and differential measurement of
  the charge asymmetry in $t\bar{t}$ events at 13 TeV with the ATLAS
  detector},'' Tech. Rep. ATLAS-CONF-2019-026, 7, 2019.

\bibitem{Ellis:2020unq}
J.~Ellis, M.~Madigan, K.~Mimasu, V.~Sanz, and T.~You, ``{Top, Higgs, Diboson
  and Electroweak Fit to the Standard Model Effective Field Theory},''
  \href{http://dx.doi.org/10.1007/JHEP04(2021)279}{{\em JHEP} {\bfseries 04}
  (2021) 279}, \href{http://arxiv.org/abs/2012.02779}{{\ttfamily
  arXiv:2012.02779 [hep-ph]}}.

\bibitem{Ethier:2021bye}
{\bfseries SMEFiT} Collaboration, J.~J. Ethier, G.~Magni, F.~Maltoni,
  L.~Mantani, E.~R. Nocera, J.~Rojo, E.~Slade, E.~Vryonidou, and C.~Zhang,
  ``{Combined SMEFT interpretation of Higgs, diboson, and top quark data from
  the LHC},'' \href{http://dx.doi.org/10.1007/JHEP11(2021)089}{{\em JHEP}
  {\bfseries 11} (2021) 089}, \href{http://arxiv.org/abs/2105.00006}{{\ttfamily
  arXiv:2105.00006 [hep-ph]}}.

\bibitem{Ellis:2018gqa}
J.~Ellis, C.~W. Murphy, V.~Sanz, and T.~You, ``{Updated Global SMEFT Fit to
  Higgs, Diboson and Electroweak Data},''
  \href{http://dx.doi.org/10.1007/JHEP06(2018)146}{{\em JHEP} {\bfseries 06}
  (2018) 146}, \href{http://arxiv.org/abs/1803.03252}{{\ttfamily
  arXiv:1803.03252 [hep-ph]}}.

\bibitem{Biekotter:2018rhp}
A.~Biekoetter, T.~Corbett, and T.~Plehn, ``{The Gauge-Higgs Legacy of the LHC
  Run II},'' \href{http://dx.doi.org/10.21468/SciPostPhys.6.6.064}{{\em SciPost
  Phys.} {\bfseries 6} no.~6, (2019) 064},
  \href{http://arxiv.org/abs/1812.07587}{{\ttfamily arXiv:1812.07587
  [hep-ph]}}.

\bibitem{Brivio:2019ius}
I.~Brivio, S.~Bruggisser, F.~Maltoni, R.~Moutafis, T.~Plehn, E.~Vryonidou,
  S.~Westhoff, and C.~Zhang, ``{O new physics, where art thou? A global search
  in the top sector},'' \href{http://dx.doi.org/10.1007/JHEP02(2020)131}{{\em
  JHEP} {\bfseries 02} (2020) 131},
  \href{http://arxiv.org/abs/1910.03606}{{\ttfamily arXiv:1910.03606
  [hep-ph]}}.

\bibitem{Hartland:2019bjb}
N.~P. Hartland, F.~Maltoni, E.~R. Nocera, J.~Rojo, E.~Slade, E.~Vryonidou, and
  C.~Zhang, ``{A Monte Carlo global analysis of the Standard Model Effective
  Field Theory: the top quark sector},''
  \href{http://dx.doi.org/10.1007/JHEP04(2019)100}{{\em JHEP} {\bfseries 04}
  (2019) 100}, \href{http://arxiv.org/abs/1901.05965}{{\ttfamily
  arXiv:1901.05965 [hep-ph]}}.

\bibitem{Buckley:2015lku}
A.~Buckley, C.~Englert, J.~Ferrando, D.~J. Miller, L.~Moore, M.~Russell, and
  C.~D. White, ``{Constraining top quark effective theory in the LHC Run II
  era},'' \href{http://dx.doi.org/10.1007/JHEP04(2016)015}{{\em JHEP}
  {\bfseries 04} (2016) 015}, \href{http://arxiv.org/abs/1512.03360}{{\ttfamily
  arXiv:1512.03360 [hep-ph]}}.

\bibitem{Grzadkowski:2010es}
B.~Grzadkowski, M.~Iskrzynski, M.~Misiak, and J.~Rosiek, ``{Dimension-Six Terms
  in the Standard Model Lagrangian},''
  \href{http://dx.doi.org/10.1007/JHEP10(2010)085}{{\em JHEP} {\bfseries 10}
  (2010) 085}, \href{http://arxiv.org/abs/1008.4884}{{\ttfamily arXiv:1008.4884
  [hep-ph]}}.

\bibitem{AguilarSaavedra:2018nen}
D.~Barducci {\em et~al.}, ``{Interpreting top-quark LHC measurements in the
  standard-model effective field theory},''
  \href{http://arxiv.org/abs/1802.07237}{{\ttfamily arXiv:1802.07237
  [hep-ph]}}.

\bibitem{Brivio:2017btx}
I.~Brivio, Y.~Jiang, and M.~Trott, ``{The SMEFTsim package, theory and
  tools},'' \href{http://dx.doi.org/10.1007/JHEP12(2017)070}{{\em JHEP}
  {\bfseries 12} (2017) 070}, \href{http://arxiv.org/abs/1709.06492}{{\ttfamily
  arXiv:1709.06492 [hep-ph]}}.

\bibitem{Degrande:2020evl}
C.~Degrande, G.~Durieux, F.~Maltoni, K.~Mimasu, E.~Vryonidou, and C.~Zhang,
  ``{Automated one-loop computations in the SMEFT},''
  \href{http://arxiv.org/abs/2008.11743}{{\ttfamily arXiv:2008.11743
  [hep-ph]}}.

\bibitem{Aad:2019mbh}
{\bfseries ATLAS} Collaboration, G.~Aad {\em et~al.}, ``{Combined measurements
  of Higgs boson production and decay using up to $80$ fb$^{-1}$ of
  proton-proton collision data at $\sqrt{s}=$ 13 TeV collected with the ATLAS
  experiment},'' \href{http://dx.doi.org/10.1103/PhysRevD.101.012002}{{\em
  Phys. Rev. D} {\bfseries 101} no.~1, (2020) 012002},
  \href{http://arxiv.org/abs/1909.02845}{{\ttfamily arXiv:1909.02845
  [hep-ex]}}.

\bibitem{Aad:2020jvx}
{\bfseries CMS, ATLAS} Collaboration, G.~Aad {\em et~al.}, ``{Combination of
  the W boson polarization measurements in top quark decays using ATLAS and CMS
  data at $\sqrt{s} =$ 8 TeV},''
  \href{http://dx.doi.org/10.1007/JHEP08(2020)051}{{\em JHEP} {\bfseries 08}
  no.~08, (2020) 051}, \href{http://arxiv.org/abs/2005.03799}{{\ttfamily
  arXiv:2005.03799 [hep-ex]}}.

\end{thebibliography}\endgroup

\end{document}